\documentclass[a4paper,reqno]{amsart}
\usepackage[text={5.0in,8in},centering]{geometry}
\usepackage[backrefs]{amsrefs}
\usepackage{amssymb}

\usepackage[usenames,dvipsnames,svgnames,table]{xcolor}
\definecolor{citegreen}{rgb}{0.00,0.70,0.30}
\usepackage[colorlinks=true,
            linkcolor=blue,
            urlcolor=blue,
            citecolor=blue]{hyperref}

\usepackage{mathrsfs}

\usepackage{xspace}
\usepackage{mathtools}

\usepackage{enumitem}

\usepackage{color}
\definecolor{trustcolor}{rgb}{0.71,0.14,0.07}

\usepackage{hyperref}

\usepackage{pgf}
\usepackage{tikz}

\usetikzlibrary{arrows,automata}

\usetikzlibrary{calc}

\DeclareMathAlphabet{\mathpzc}{OT1}{pzc}{m}{it}

\usepackage{epsfig}
\usepackage{graphicx}

\numberwithin{equation}{section}
\theoremstyle{plain}
\newtheorem{theorem}{Theorem}
\newtheorem{prop}[theorem]{Proposition}

\newtheorem{lemma}{Lemma}

\newtheorem{corollary}[theorem]{Corollary}
\theoremstyle{remark}
\newtheorem{remark}{Remark}[section]

\newtheorem*{quest*}{Question}
\newtheorem*{remark*}{Remark}
\theoremstyle{remark}

\theoremstyle{definition}
\newtheorem{definition}{Definition}[section]
\newtheorem*{definition*}{Definition}
\newtheorem*{notation*}{Notation}
\newtheorem*{notations*}{Notations}
\providecommand{\B}{\mathbf}

\providecommand{\D}{\mathbb}

\newcommand{\ee}{\mathrm{e}}
\newcommand{\eu}{\mathrm{e}}

\DeclareMathOperator{\Ord}{O}
\DeclareMathOperator{\sord}{o}

\DeclareMathOperator{\tr}{tr\,}

\DeclareMathOperator*{\essup}{ess\,sup}
\DeclareMathOperator*{\supp}{supp}

\DeclareMathOperator{\one}{\mathbf{1}}

\DeclareMathOperator{\Unif}{Unif}

\def\mytimes{\operatornamewithlimits{\hbox{\huge$\times$}}}
\def\mytimesL{\operatornamewithlimits{\hbox{\LARGE$\times$}}}

\def\lam{{\lambda}}

\def\a{\alpha}

\def\eps{\epsilon}

\def\prk#1{\mathbb{P}_{\Bk}\left\{ #1 \right\}}

\def\esmk#1{\D{E}_{\Bk}\left[\, #1\, \right]}

\def\rhob{\overline{\rho}}

\def\cond{\,\big|\,}

\def\BP{\B{P}}

\def\tBPk{\widetilde{\B{P}}_{\Bk}}

\def\BC{\mathbf{C}}

\def\BJ{\mathbf{J}}

\def\BJell{\BJ^{(\ell)}}

\def\BK{\mathbf{K}}

\def\BX{\mathbf{X}}

\def\Ba{\mathbf{a}}

\def\Bk{\mathbf{k}}
\def\Bp{\mathbf{p}}

\def\tBp{\widetilde{\mathbf{p}}}

\def\Bx{\mathbf{x}}
\def\By{\mathbf{y}}

\def\tileta{\tilde{\eta}}
\def\txi{\tilde{\xi}}
\def\tI{\tilde{I}}
\def\teta{\tilde{\eta}}
\def\tmu{\tilde{\mu}}
\def\tlam{\tilde{\lam}}

\def\DP{\D{P}}
\def\DR{\D{R}}
\def\DY{\D{Y}}
\def\DZ{\D{Z}}

\def\cA{\mathcal{A}}

\def\cH{\mathcal{H}}

\def\cK{\mathcal{K}}

\def\bcK{\boldsymbol{\mathcal{\cK}}}

\def\cN{\mathcal{N}}

\def\cS{\mathcal{S}}

\def\cX{\mathcal{X}}
\def\tcX{\widetilde{\mathcal{X}}}

\def\cA{{\mathcal{A}}}
\def\cN{{\mathcal{N}}}

\def\uX{\underline{X}}

\def\oX{\overline{X}}

\def\be{\begin{equation}}
\def\bel#1{\begin{equation}\label{#1}}
\def\ee{\end{equation}}

\def\ba{\begin{array}{l}}
\def\ea{\end{array}}
\def\bal{\begin{aligned}}
\def\eal{\end{aligned}}

\def\fF{\mathfrak{F}}
\def\fFbcK{\mathfrak{F}_{\boldsymbol{\mathcal{K}}}}

\def\fl{\mathfrak{l}}

\def\om{{\omega}}
\def\Om{{\Omega}}

\def\eps{\epsilon}

\def\Lam{{\Lambda}}
\def\lam{{\lambda}}

\def\pr#1{\D{P}\left\{\,#1\,\right\}}
\def\esm#1{\D{E}\left[\, #1\, \right]}

\def\half{\frac{1}{2}}

\def\ble{\begin{lemma}}
\def\ele{\end{lemma}}

\def\bre{\begin{remark}}
\def\ere{\end{remark}}

\def\btm{\begin{theorem}}
\def\etm{\end{theorem}}

\def\bde{\begin{definition}}
\def\ede{\end{definition}}

\def\bpr{\begin{prop}}
\def\epr{\end{prop}}

\def\bco{\begin{corollary}}
\def\eco{\end{corollary}}

\begin{document}

\title[On the conditional distribution of the sample mean]
{Optimized estimates of the regularity\\ of the conditional distribution\\of the sample mean}

\author[V. Chulaevsky]{Victor Chulaevsky}


\address{D\'{e}partement de Math\'{e}matiques\\
Universit\'{e} de Reims, Moulin de la Housse, B.P. 1039\\
51687 Reims Cedex 2, France\\
E-mail: victor.tchoulaevski@univ-reims.fr}

\date{}
\begin{abstract}
We give an improved estimate for the regularity of the conditional distribution
of the empiric mean of a finite sample of IID random variables,
conditional on the sample "fluctuations", extending the well-known property of Gaussian IID samples.
Specifically, we replace the bounds in probability, established in our earlier works, by those in distribution, and this
results in the optimal regularity exponent in the final estimate.
\end{abstract}

\maketitle

\section{Introduction} \label{sec:intro}

Consider a sample of $N$ IID (independent and identically distributed)
random variables with Gaussian distribution $\cN(0,1)$, and
introduce the sample mean $\xi=\xi_N$ and the "fluctuations" $\eta_i$ around the mean:
$$
\xi_N = \frac{1}{N} \sum_{i=1}^N X_i,
\quad \eta_i = X_i - \xi_N, \;\; i=1, \ldots, N.
$$
It is well-known from elementary courses of the probability theory that $\xi_N$ is independent
from the sigma-algebra $\fF_\eta$ generated by $\{\eta_1, \ldots, \eta_n\}$ (the latter
are linearly dependent, and have rank $N-1$). To see this, it suffices
to note that $\eta_i$  are all orthogonal to $\xi_N$ with respect to the standard
scalar product in the linear space formed by $X_1, \ldots, X_N$ given by
$$
\langle Y, Z \rangle := \esm{ Y \, Z} ,
$$
where $Y$ and $Z$ are real linear combinations of $X_1, \ldots, X_N$ (recall: $\esm{X_i}=0$).

Therefore, the conditional probability distribution of $\xi_N$ given $\fF_\eta$ coincides
with the unconditional one, so $\xi_N \sim \cN(0, N^{-1})$, thus $\xi_N$ has bounded density
$$
p_\xi(t) = \frac{e^{ -\half t^2} }{ \sqrt{2\pi N^{-1}} } \le \frac{N^{1/2}}{ \sqrt{2\pi}}.
$$
Moreover, for any interval $I\subset\DR$ of length $|I|$, we have
\be\label{eq:Gaussian.SRCM}
\essup \pr{ \xi_N(\om) \in I \,\big|\, \fF} = \pr{ \xi_N(\om) \in I }
\le \frac{N^{1/2}}{ \sqrt{2\pi}} \, |I|.
\ee
The essential supremum in the above LHS is a  bureaucratic tribute to the formal
rule saying that $\pr{\,\cdot\, \,|\, \fF}$ is a random variable (which is $\fF$-measurable),
and as such is defined, generally speaking, only up to subsets of measure zero.

In some applications to the eigenvalue concentration estimates in the theory of multi-particle
random, Anderson-type Hamiltonians, one has to estimate
the probability of the form
$$
\pr{ \xi_N(\om) \in I(\eta) } ,
$$
where the interval $I(\eta) =[f(\eta), f(\eta)+\eps]$ is determined only by the fluctuations $\eta_\bullet$,
and $f$ is some measurable (in fact, Lipschitz
continuous\footnote{We refer to the applications where $f$ is an eigenvalue of some self-adjoint operator,
and by the min-max principle, such EVs are Lipschitz continuous functions of the parameters
upon which the operator depends.})
function. For example, with $N=2$,
$$
\xi = \xi_2 = \frac{X_1 + X_2}{2}, \;\; \eta = \eta_1 = \frac{X_1 - X_2}{2},
$$
one may consider the probability
$$
\pr{ \xi \in [\eta^2, \eta^2 + s] }
= (2\pi)^{-1} \int_{\DR^2} dX_1\, dX_2 \, \eu^{-\half(x_1^2 + x_2^2) }
\one_A(x_1,x_2)
$$
where, e.g.,
$$
A := \left\{(x_1,x_2)\in\DR^2:\;  \frac{(x_1 - x_2)^2}{4} \le \frac{x_1 + x_2}{2} \le \frac{(x_1 - x_2)^2}{4} + s  \right\}, \; s>0.
$$

\begin{center}
\begin{figure}

\begin{tikzpicture}
\begin{scope}[scale=0.75]

\draw[->,color=black, line width = 1.0] (-0.9, -0.7) -- (3.9, -0.7);
\draw[->,color=black, line width = 1.0] (-0.7, -0.9) -- (-0.7, 3.3);

\node at (3.9, 3.4) (xi) {$\xi$};
\node at (1.3, -2.1) (eta) {$\eta$};

\begin{scope}
\clip (-1.0,-1.0) rectangle ++(4.0, 4.0);

\foreach \a in {0.25, 0.20, 0.15, 0.10, 0.05}
\draw[color=red!20!white!80, , line width = 2.2, samples at={-0.048, 0.002, ..., 4.0}]
plot(\x, {-\a + 0.5/(\x + \a)});

\foreach \a in {0, 2, ..., 10}
\draw[color=gray] (-2+0.2*\a, -1) -- (-2 + 0.2*\a + 5, -1+5);

\draw[color=red, line width = 2.0] (0.1, 1.1) -- (0.1+ 0.28,  1.1 + 0.28);

\draw[color=red, line width = 2.0] (0.20 , 0.8) -- (0.20 + 0.28,  0.8 + 0.28);

\draw[color=red, line width = 2.0] (0.34 , 0.55) -- (0.34 + 0.28,  0.55 + 0.28);

\draw[color=red, line width = 2.0] (0.53 , 0.35) -- (0.53 + 0.28,  0.35 + 0.28);

\draw[color=red, line width = 2.0] (0.80 , 0.22) -- (0.80 + 0.28,  0.22 + 0.28);

\draw[color=red, line width = 2.0] (1.10 , 0.12) -- (1.10 + 0.27,  0.12 + 0.27);

\draw[color=red!50!black!80, line width = 1.5, samples at={0.12, 0.17, ..., 4.0}]
plot(\x, {0.5/\x });

\draw[color=red!50!black!80, line width = 1.4, samples at={-0.102, -0.055, ..., 4.0}]
plot(\x, {-0.25 + 0.5/(\x+0.25)});

\end{scope}

\draw[->,color=blue, line width = 1.0] (-2.5, 1.1) -- (1.1, -2.5);
\draw[->,color=blue, line width = 1.0] (-0.9, -0.9) -- (3.5, 3.5);

\end{scope}
\end{tikzpicture}

\caption{In this example, $N=2$, $\xi = \half(X_1 + X_2)$ and $\eta = \half(X_1 - X_2)$.
One has to assess the probability of the pink curvilinear strip $\{(X_1,X_2):\, \xi\in[a(\eta), a(\eta)+s\}$ .}

\end{figure}
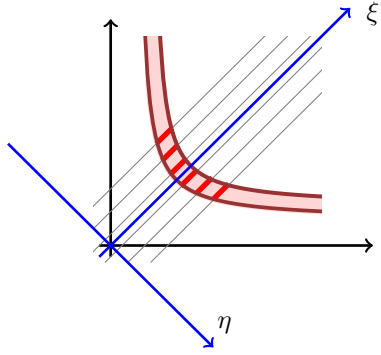
\end{center}

In this particular case -- for Gaussian samples -- the conditional regularity of the sample
mean $\xi_N$ (given the fluctuations) is granted, but is not always so, as shows the following
elementary example where the common probability distribution of the sample $X_1, X_2$
is just excellent: $X_i \sim \Unif([0,1])$, so $X_i$ admit a compactly supported
probability density bounded by $1$. In this simple example
the random vector $(X_1,X_2)$ is uniformly distributed in the unit square $[0,1]^2$,
and the condition $\eta = c$ selects a straight line in the two-dimensional plane
with coordinates $(X_1,X_2)$, parallel to the main diagonal $\{X_1 = X_2\}$. The conditional distribution
of $\xi$ given $\{\eta = c\}$ is the uniform distribution on the segment
$$
J_c := \{(x_1, x_2):\; x_1-x_2 = 2c,\, 0\le x_1, x_2 \le 1\}
$$
of length vanishing at $2c = \pm 1$. For $|2c|=1$, the conditional distribution of $\xi$
on $J_c$ is concentrated on a single point, which is the ultimate form of singularity.

\section{An application to the Wegner-type bounds}
\label{sec:applic.to.MSA}

Let $\Lam$ be a finite graph, with $|\Lam|=N\ge 1$,
and $H(\om)=H_\Lam(\om)$ be a random DSO acting in the finite-dimensional
Hilbert space $\cH = \cH_\Lam = \ell^2(\Lam)$, with IID random potential potential $V:\Lam\times\Om\to\DR$,
relative to a probability space $(\Om,\fF,\DP)$. Decomposing the random field $V$ on $\Lam$,
$$
V(x;\om) = \xi_N(\om) + \eta_x(\om),
$$
we can represent $H(\om)$ as follows:
$$
H(\om) = \xi_N(\om) \one + A(\om),
$$
where the self-adjoint operator $A(\om)$ is $\fF_\eta$-measurable, and so are its eigenvalues
$\tmu_j(\om)$, $j=1, \ldots, N$. It is readily seen that $A(\om)$ is a DSO with potential having zero
sample mean.
Since $A(\om)$ commutes with the scalar operator $\xi_N(\om)\one$,
the eigenvalues $\lam_j(\om)$ of $H(\om)$ have the form
\be\label{eq:lam.xi.tmu}
\lam_j(\om) = \xi_N(\om) + \mu_j(\om) .
\ee
The numeration of the eigenvalues $\lam_j(\om)$, $\mu_j(\om)$ is, of course, not canonical, but
they can be consistently defined as random variables on $\Om$.

The representation \eqref{eq:lam.xi.tmu} implies immediately the following EVC bound:
for any interval $I_s = [t, t+s]$,
\be
\bal
\pr{ \tr P_{I_s}(H(\om)) \ge 1} &\le \sum_{j=1}^N \pr{ \lam_j(\om) \in I_s}
= \sum_{j=1}^N \pr{ \xi_N(\om) + \mu_j(\om)  \in I_s}
\\
& = \sum_{j=1}^N \esm{ \pr{ \xi_N(\om) + \mu_j(\om)  \in I_s \cond \fF_\eta} }
\\
& = \sum_{j=1}^N \esm{ \pr{ \xi_N(\om) \in [- \mu_j(\om) + t, - \mu_j(\om) + t+s]\cond \fF_\eta} }
\eal
\ee
Further, omitting the argument $\om$ for notational brevity, we have
$$
\bal
\pr{ \xi_N + \tmu_j  \in I_s \cond \fF_\eta}
&= \pr{ \xi_N \in [ \mu_j + t,  \mu_j + t+s]\cond \fF_\eta}
\\
& = \pr{ \xi_N \in [\tmu_j, \tmu_j +s]\cond \fF_\eta}
\eal
$$
where $\tmu_j(\om) := -\mu_j(\om) + t$ are $\fF_\eta$-measurable, i.e., fixed under the
conditioning. Now introduce the conditional continuity modulus of $\xi_N$, given $\fF_\eta$:
$$
\nu_N(s) := \sup_{t\in\DR} \; \essup \; \pr{ \xi_N \in [t, t+s]\cond \fF_\eta}, \;\; s>0.
$$
Obviously,
$$
\pr{ \lam_j  \in I_s \cond \fF_\eta} \le \nu_N(s),
$$
thus the unconditional probability $\pr{ \lam_j  \in I_s }$ can be assessed by analyzing the
probability distribution of the random conditional continuity modulus $\nu_N(s;\om)$.

In this section, we discuss by way of example the Wegner-type bounds for a conventional,
single-particle DSO, but in applications to the multi-particle EVC bounds, similar objects
turn out to be of interest:
\be\label{eq:general.nu.mu.bound}
s \mapsto \pr{ \xi_N(\om) \in [\tmu(\om), \tmu(\om) + s  },
\ee
where an $\fF_\eta$-measurable random variable $\tmu$ is given by an eigenvalue of yet another
operator $\widetilde{H}(\om)$ which is not necessarily independent of $H(\om)$.
The most difficult case is where $H(\om)$ and $\widetilde{H}(\om)$ are stochastically correlated
in a very strong way: every "local" random variable, representing the disorder in a multi-particle Anderson model,
which affects $H(\om)$ also affects $\widetilde{H}(\om)$, and vice versa. As a result, there is little one
can say about $\tmu(\om)$, except that it is a measurable function.

\section{Reduction to the local analysis in the sample space}
\label{sec:partition}

Assume that the support $\cS\subset \DR$
of the common \emph{continuous} marginal probability measure $\DP_V$ of the IID random variables
$X_j$, $1\le j \le N$, is covered by a finite or countable union of  intervals:
$$
\cS \subset \cup_{k\in \cK} J_k, \;\;\cK\subset\DZ, \; J_k = [a_k, b_k], \;\; a_{k+1} \ge b_k.
$$
Let $\BK = \cK^N$, and for each $\Bk = (k_1, \ldots, k_N)\in\BK$, denote
$$
\BJ_\Bk = \mytimesL_{i=1}^N J_{k_i}.
$$
Owing to the continuity of the marginal measure, $J_k$ are "essentially" disjoint:
for all $k\ne l$, $\DP_V(J_k \cap J_l) = 0$. Respectively, the family of the parallelepipeds
$\{\BJ_\Bk, \; \Bk\in\BK\}$ forms a partition $\bcK$ of the sample space, which we will often identify
with the probability space $\Om$. Further, let $\fFbcK$ be the sub-sigma-algebra of $\fF$
generated by the partition $\bcK$. Now the quantities of the general form
\eqref{eq:general.nu.mu.bound} can be assessed as follows:
$$
\bal
\pr{ \xi_N \in [\tmu, \tmu +s]} & = \esm{ \pr{ \xi_N \in [\tmu, \tmu +s] \cond \fFbcK} }
\\
& = \sum_{\Bk\in\BK} \pr{ \BJ_\Bk } \pr{ \xi_N \in [\tmu, \tmu +s] \cond \BJ_\Bk }.
\eal
$$
Let $\prk{\cdot}$ be the conditional probability measure, given $\{X\in\BJ_\Bk\}$,
$\esmk{\cdot}$ the respective expectation, and $p_\Bk = \pr{ \BJ_\Bk }$. Then we have
\be\label{eq:pr.xi.mu.BK}
\bal
\pr{ \xi_N \in [\tmu, \tmu +s]}
 &= \sum_{\Bk\in\BK} p_\Bk \esmk{ \prk{ \xi_N \in [\tmu, \tmu +s] \cond \fF_\eta } }
\\
& \le \sup_{\Bk\in\BK}  \esmk{ \prk{ \xi_N \in [\tmu, \tmu +s] \cond \fF_\eta } }.
\eal
\ee
This simple formula shows that one may seek a satisfactory upper bound on the LHS of
\eqref{eq:pr.xi.mu.BK} by assessing the "local" conditional probabilities
$\prk{ \xi_N \in [\tmu, \tmu +s] \cond \fF_\eta }$, where each random variable $X_j$
is restricted to a subinterval $J_{k_j}$ of its global support, so
the entire sample $X=(X_1, \ldots, X_N)$ is restricted to a parallelepiped
$\BJ \subset\DR^N$.

In the next section, we perform such analysis first in the case of a uniform marginal distribution
of the IID variables $X_i$.

\section{Uniform marginal distributions}
\label{sec:Unif}

Let be given a real number $\ell > 0$ and an integer $N\ge 2$.
Consider a sample of $N$ IID random variables with uniform distribution
$\Unif([0,\ell])$, and
introduce again the sample mean $\xi=\xi_N$ and the "fluctuations" $\eta_i$ around the mean:
$$
\xi_N = \frac{1}{N} \sum_{i=1}^N X_i,
\quad \eta_i = X_i - \xi_N.
$$
For the purposes of orthogonal transformation
$(X_1, \ldots, X_n) \mapsto (\txi_N, \teta_2, \ldots, \teta_N)$, we also need a rescaled
empirical mean
$$
\txi_N = N^{1/2} \xi_N,
$$
so
\be\label{eq:X.i.txi.eta}
X_i = \eta_i + N^{-1/2} \txi_N , \;\; i=1, \ldots N.
\ee
Further, consider the Euclidean space $\sim \DR^N$ of real linear combinations of the random variables
$X_i$ with the scalar product $\langle X', X''\rangle = \esm{X' X''}$.
Clearly, the variables $\eta_i:\DR^N\to\DR$ are invariant under the group of translations
$$
(X_1, \ldots, X_N) \mapsto (X_1 + t, \ldots, X_N+t), \;\; t\in\DR,
$$
and so are their differences $\eta_i - \eta_j \equiv X_i - X_j$, $1\le i < j \le N$.
Introduce the variables
\be\label{eq:def.Y.i}
Y_i = \eta_i - \eta_N, \; \; 1\le i \le N-1,
\ee
Then the space $\DR^N$ is fibered into a union of affine lines of the form
\be\label{eq:def.Y.i.cX}
\bal
\tcX(Y) &:= \{X\in\DR^N:\, \eta_i - \eta_N = Y_i, \, i\le N-1\}
\\
&:= \{X\in\DR^N:\, X_i - X_N = Y_i, \, i\le N-1\},
\eal
\ee
labeled by the elements $Y = (Y_1, \ldots, Y_{N-1})$ of the $(N-1)$-dimensional real
vector space $\DY^{N-1} \cong \DR^{N-1}$.
Set
$$
\cX(Y) = \tcX(Y) \cap \BC_1
= \{X\in\BC_1:\, X_i - X_N = Y_i, \, i\le N-1\}
$$
and endow each nonempty interval $\cX(Y)\subset\DR^N$
with the natural structure of a probability space inherited from $\DR^N$:
\begin{itemize}
  \item if $|\cX(Y)|=0$ (an interval reduced to a single point), then we introduce the
  trivial sigma-algebra and trivial counting measure;
  \item if $|\cX(Y)|=r>0$, then we use the inherited structure of an interval of a
  one-dimensional affine line and the normalized measure with constant density $r^{-1}$
      with respect to the inherited Lebesgue measure on $\cX(Y)$.
\end{itemize}

The transformation $X \mapsto (\xi_N, \eta_1, \ldots, \eta_{N-1})$ is non-degenerate, but
not orthogonal. We will have to work with the metric on $\cX(Y)$, induced by the standard
Riemannian metric in the ambient space $\DR^N$; to this end, introduce an orthogonal coordinate transformation in $\DR^N$,
$X \mapsto (\txi_N, \tileta_1, \ldots, \tileta_{N-1})$, such that
\be\label{eq:txi.N}
\txi_N = N^{-1/2} \sum_{i=1}^N X_i = N^{1/2} \xi_N;
\ee
the exact form of $\tileta_j$, $j=1, \ldots, N-1$ is of no importance, provided
that the transformation is orthogonal.

\bre\label{rem:X.i.scaled.param}
For later use, note that, owing to \eqref{eq:txi.N}, each of the re-scaled variables $N^{1/2}X_i$ can
serve as the (normalized) length parameter on the elements $\cX(Y)$.
Along an element $\cX(Y)$, one can simultaneously parameterize $\txi$ and the variables $X_i$,
by setting $\txi(t) = c_0 + t$, $X_j(t) = c_j + N^{-1/2} t$,
with arbitrarily chosen constants $c_j$. Here, $\txi_N$ is a natural length parameter on $\cX(Y)$,
since the transformation $X \mapsto (\txi_N, \tileta_1, \ldots, \tileta_{N-1 })$ is orthogonal.
\ere

It follows from \eqref{eq:txi.N} that for any given $a\in\DR$, $s> 0$,
and some $a'\in\DR$,
\be\label{eq:xi.nu.eps}
\bal
\xi_N \in [a, a + s] &\Longleftrightarrow
 \txi_N \in[a', a' + N^{1/2}s]
\eal
\ee

Next, denote $\BJ^{(\ell)} = [0,\ell]^N$ and introduce the random variable
\be\label{eq:nu.s}
\bal
\nu_N(s; \BJell) = \nu_N(s;\BJell; X) &:=
\essup\; \sup_{t\in\DR} \pr{ \xi_N\in[t, t + s] \,\big|\, \fF_\eta} .
\eal
\ee
Here the presence of $\essup$ is the tribute to the fact that the conditional probabilities are random
variables, usually defined up to subsets of zero measure; $\ell>0$ is the width of the common uniform
distribution of $X_j$. Equivalently, one may write $\nu_N(s;\BJell; \om)$ instead of
$\nu_N(s;\BJell; X)$, since the sample space $\DR^N$ is identified with the underlying
probability space $\Om$.

Since $\{X_i\}$ are IID with uniform distribution on $[0,\ell]$, the distribution of the random
vector $X(\om)$ is uniform in the cube $\BJell=[0,\ell]^N$, inducing a uniform conditional distribution
on each element $\cX(Y)$. Therefore, by \eqref{eq:xi.nu.eps} and \eqref{eq:nu.s},
\be\label{eq:nu.eps.1}
\bal
\nu_N(s;\BJell) = \frac{ N^{1/2} s}{ |\cX(Y) | } .
\eal
\ee

It is to be stressed that both sides of the above equality are random variables:
$\nu_N(s;\ell) = \nu_N(s;\ell;\om)$ by its definition in \eqref{eq:nu.s}, and
$\cX(Y) = \cX(Y(X(\om)))$.

\section{Short intervals are unlikely}

\ble\label{lem:prob.small.cX.density}
Assume that the IID random variables $X_1, \ldots, X_N$, $N\ge 2$, admit (common) probability density
$p_V$ with $\|p_V\|_\infty \le \rhob<\infty$. Then
\be\label{eq:lem.small.cX.density.1}
\pr{ |\cX(Y)| < r } \le \frac{1}{4} \rhob^2 r^2 N.
\ee
In particular, for $X_j \sim \Unif([0,\ell))$, one has
\be\label{eq:lem.small.cX.density.2}
\pr{ |\cX(Y)| < r } \le \frac{r^2 N}{4 \ell^2}.
\ee
\ele
\proof

Let
\be\label{eq:def.X.star}
\uX = \uX(X) = \min_{i} X_i, \; \oX = \oX(X)=\max_{i} X_i.
\ee
While $\oX(X)$ and $\uX(X)$ vary along the elements $\cX(Y)$, their difference
$\oX(X) -\uX(X)$ does not; it is uniquely determined by $\cX(Y)$.

According to Remark \ref{rem:X.i.scaled.param}, each $N^{1/2}X_i$, $i=1, \ldots, N$,
restricted to $\cX(Y)$, provides a normalized length parameter on $\cX(Y)$; thus the range
of each $N^{1/2} X_i |_{\cX(Y)}$ is an interval of length $|\cX(Y)|$. One can increase (resp., decrease),
e.g., the value of $X_1$, as long as \emph{all} $\{X_i, 1\le i \le N\}$ are strictly smaller than $\ell$
(resp., strictly positive). Therefore, the maximum increment of $X_1$
(indeed, of any $X_i$) along $\cX(Y)$
is given by $\ell - \oX(X)$, and its maximum decrement equals $\uX(X)$, so the range
of the normalized length parameter $N^{1/2} X_1$ along $\cX(Y(X))$ is
an interval of length $N^{1/2}\big(\ell - \oX(X) + \uX(X) \big)$:
\be\label{eq:len.cX.oX.uX.again}
|\cX(Y(X))| = N^{1/2} \big( \ell - \oX(X) + \uX(X) \big),
\ee
Since both $\uX(X)$ and $\ell -\oX(X)$ are non-negative,
\be\label{eq:def.r.Y}
 \uX + (\ell -\oX) < t  \;\; \Longrightarrow \;\; \max\{\uX,\; \ell - \oX\} < t.
\ee
With $0 \le t\le \ell$, $\big(\ell-X_i < t/2\big)$ implies $\big(X_i > t/2\big)$, thus
denoting
\be\label{eq:A.i.i.empty.1}
 A_{ij}(t) := \{  X_i < t/2 \} \cap \{ \ell-X_j < t \},
\ee
we have, for any $i$,
\be\label{eq:A.i.i.empty.2}
A_{ii}(t) = \{ X_i < t \}  \, \cap \, \{ \ell-X_i < t \} = \varnothing.
\ee
Therefore,
\be\label{eq:sum.Akij.unif}
\left\{\max\big\{\uX(X),\; \ell-\oX(X)\big\} < t  \right\} \subset
\bigcup_{i \ne j} \left\{  X_i < \frac{t}{2}, \; \ell-X_j < t \right\}.
\ee
Thus the union $\cup_{i\ne j} A_{ij}(t)$ contains all samples $X$ with $|\cX(Y)| < t$.

The sample $\{X_k\}$ is IID, with common probability density uniformly bounded by $\rhob<\infty$,
so for any $i\ne j$
$$
\pr{A_{ij}(t)} = \pr{ X_i < t } \cdot \pr{ \ell-X_j < t }
=  \rhob^2 t^2.
$$
Therefore,
\be\label{eq:prob.Aij.to.r}
\bal
\pr{ |\cX(Y)| < r} &= \pr{ N^{1/2} \big( (\ell - \oX(X)) + \uX(X) \big) < r }
\\
& = \pr{ \big( (\ell - \oX(X)) + \uX(X) \big) < r N^{-1/2} }
\\
&
\le \sum_{i \ne j} \pr{ A_{ij}\big( r N^{-1/2} \big) }
  \le  N(N-1) \,  \left(\rhob  r N^{-1/2}\right)^2
\\
& \le  \rhob^2 r^2 N.
\eal
\ee

\qedhere

\section{Regularity bound for the uniform distributions}

\btm\label{thm:main.unif}
Let be given IID random variables $X_1, \ldots, X_N$ with $X_i \sim \Unif([0,\ell])$
and a measurable function $\lam:\, Y \mapsto \lam(Y)$. In each interval $\cX(Y)\subset \tcX(Y) \cong \DR$,
introduce the sub-interval
$I_s(Y)$ $= [\lam(Y), \lam(Y)+s]\cap\tcX(Y)$.
For any $s \in(0,1]$,
\be\label{eq:thm.nu.xi.unif.ell.1}
\bal
\pr{ \xi(\om) \in I_s(Y) } &
\le \frac{3 N^{3}}{\ell} s~.
\eal
\ee
\etm

\proof
Let $\fl(\om) := |\cX(Y)|$. The function $\xi$ cannot serve as a normalized length parameter
on the intervals parallel to $(1, \ldots, 1)$, since its gradient 
$(1/N, \ldots, 1/N)$ has norm $1/\sqrt{N}$. For this reason, it is convenient to introduce
its normalized counterpart $\txi = \xi \sqrt{N}$ and rescaled intervals 
$\tI_s = [\tlam, \tlam + s\sqrt{N}]$, $\tlam = \lam \sqrt{N}$.

\be\label{eq:prob.xi.in.I.s}
\bal
\pr{ \xi\in I_s(\eta) } &= \pr{ \txi\in \tI_s(\eta) } = \esm{ \pr{ \txi\in \tI_s(\eta) \cond \fF_\eta } }
\\
& = \esm{ \one_{ \fl(\om) < s \sqrt{N}} \pr{ \txi\in \tI_s(\eta) \cond \fF_\eta } } 
           + \esm{ \one_{ \fl(\om) \ge s \sqrt{N}} \pr{ \txi\in \tI_\eps(\eta) \cond \fF_\eta } }
\\
& \le \pr{ \fl(\om) < s\sqrt{N}} 
  + \esm{ \one_{ \fl(\om) \ge s\sqrt{N}} \pr{ \txi\in \tI_\eps(\eta) \cond \fF_\eta } }
\eal
\ee
where, by virtue of \eqref{eq:prob.Aij.to.r},
\be\label{eq:prob.ell.le.s}
\pr{ \fl(\om) < s\sqrt{N}} \le \frac{N^2}{ \ell^2} s^2,
\ee
yielding
\be\label{eq:ratio.F.ell.l.2}
\sup_{s>0}
\frac{\pr{ \fl(\om) < s }}{s^2}  \le \frac{N^2}{ \ell^2}.
\ee
The second summand in the RHS of \eqref{eq:prob.xi.in.I.s} can be assessed as follows:
\be\label{eq:expectation.decomp}
\bal
\esm{ \one_{ \fl \ge s \sqrt{N}} \pr{ \txi\in \tI_s(\eta) \cond \fF_\eta } }
&\le \esm{ \one_{ \fl \ge s\sqrt{N}} \frac{s\sqrt{N} }{ \fl} }
= s\sqrt{N}\, \esm{ \one_{ \fl \ge s} \fl^{-1} }
\\
&= s\sqrt{N}\, \int_{s\sqrt{N}}^{\ell\sqrt{N}} r^{-1} \, dF_\fl(r)
\eal
\ee
\noindent
Using integration by parts for the Stiltjes integral and \eqref{eq:ratio.F.ell.l.2}, we obtain
\be\label{eq:Stiltjes.by.parts}
\bal
\int_{s\sqrt{N}}^{\ell\sqrt{N}} r^{-1} \, dF_\fl(r) 
   & = \frac{F(r)}{r}\Big|_{s\sqrt{N}}^{\ell\sqrt{N}} 
       + \int_{s\sqrt{N}}^{\ell\sqrt{N}} r^{-2} \, F_\fl(r)\, dr
\\
& \le \frac{1}{\ell \sqrt{N}} + \ell\sqrt{N} \sup_{r>0} \frac{F_\fl(r)}{r^2}
\le \frac{1}{\ell \sqrt{N}} + \frac{\ell\sqrt{N}\cdot N^2}{ \ell^2}
\\
& \le \frac{2 N^{5/2}}{\ell}~.
\eal
\ee
Collecting \eqref{eq:prob.ell.le.s}, \eqref{eq:expectation.decomp} and \eqref{eq:Stiltjes.by.parts}, 
and taking into account that $s/\ell \le 1$, the assertion follows:
\be\label{eq:main.unif.final.bound}
\bal
\pr{ \xi\in I_s(\eta) }  \le  \frac{N^2}{\ell^2} s^2 + \frac{2N^{5/2}}{\ell} s
\le \frac{3 N^{3}}{\ell} s~.
\eal
\ee
\qedhere

\section{Smooth positive densities }
\label{sec:smooth}

Now we consider a richer class of probability distributions. While the conditions
which we will assume are certainly very restrictive, they are quite sufficient
for applications to physically realistic Anderson models.

\btm\label{thm:nu.xi.ell.piecewise}
Assume that the common probability distribution of the IID random variables
$V_j, \, j=1, \ldots, N$,
with  PDF $F_V$,  satisfies the following conditions:
\begin{enumerate}[label=\rm(\roman*),leftmargin=2.1em,align=right]
  \item the probability distribution is absolutely continuous:
\be\label{eq:cond.rho.support}
dF_V(v) = \rho(v)\, dv, \; \supp \rho = [a, a+\ell];
\ee
  \item the probability density $\rho(\cdot)$  has bounded logarithmic derivative on $(a, a+\ell)$:
\be\label{eq:cond.rho.log.deriv}
 \left\| (\ln \rho)' \, {\one_{(a, a+\ell)}}\right\|_\infty \le C'_{\rho} < +\infty.
\ee
\end{enumerate}
Then there exists a constant $C = C(F_V,\ell)<\infty$ such that for any $s \in(0, \ell N^{-2})$ and any $\fF_\eta$-measurable random variable
$\lam$, setting $I_s(\om) := [\lam(\om), \lam(\om)+s]$, one has the following bound:
\be\label{eq:thm.xi.smooth}
\bal
\pr{ \xi_N(\om) \in I_s(\om) } \le C N s.
\eal
\ee
\etm

\proof
Without loss of generality, it suffices to prove the claim for $\supp \rho =[0, \ell]$, which we assume below.

\par\vskip2mm\noindent
$\blacklozenge$
As in Section \ref{sec:partition}, introduce a partition of the sample space into the cubes $\BJ_\Bk $,
induced by the decomposition $[0, \ell] = \sqcup_k J_k$,
$$
J_k = \left[ \frac{k-1}{M_N}, \frac{k}{M_N} \right], \;\; k=1, \ldots, M_N = N^2.
$$
We have then
$$
\BJ_\Bk = \mytimes_{i=1}^N J_{k_i}, \;\; \Bk = (k_1, \ldots, k_N).
$$

\par\vskip2mm\noindent
$\blacklozenge$
The hypothesis \eqref{eq:cond.rho.log.deriv} implies that for any $\Bx\in\BJ_\Bk$ the logarithm of $\Bp(\Bx)$
is well-defined and satisfies
$$
\bal
| \ln \Bp(\Bx) - \ln \Bp(\Ba_\Bk) | \le \sum_{i=1}^N |\ln \rho(x_i) - \ln \rho(a_{k_i})|
\le N \, C'_p \, \ell M_N^{-1} = \Ord( \ell N^{-1}).
\eal
$$
thus, setting $\alpha_N = \ell N^{-1}$,
$$
\forall\, \Bx\in\BJ_\BK\quad \frac{ \Bp(\Bx)}{ \Bp(\Ba_\Bk)} \in \left[ \eu^{-\alpha_N}, \eu^{+\alpha_N}  \right].
$$

Now introduce in $\BJ_\Bk$:
\begin{itemize}
  \item the uniform probability distribution $\tBPk$, i.e., the normalized measure with constant density $\tBp_\Bk$ w.r.t. the Lebesgue measure;

  \item the probability distribution induced by $\BP$, conditional on $\{\BX\in\BJ_\Bk\}$, i.e., the normalized measure
with density
$$
\Bp_\Bk(\Bx) = Z^{-1}_{\Bk} \Bp(\Bx) = \frac{\Bp(\Bx)}{ \int_{\BJ_\Bk} \BP(\By)\, d\By}
$$
\end{itemize}
By continuity of the density $\Bp$, $ \int_{\BJ_\Bk} \BP(\By)\, d\By = c|\BJ_\Bk|$,
for some $c\in\left[ \eu^{-\alpha_N}, \eu^{+\alpha_N}\right]$, so
$$
\frac{\Bp_\Bk(\Bx)}{ \tBp(\Bx)} = \frac{\Bp(\Bx)}{c } \in\left[ \eu^{-2\alpha_N}, \eu^{+2\alpha_N}  \right]
$$
Hence for any event $\cA$, we have
\be\label{eq:prob.BJ.alpha}
\eu^{-2\alpha_N} \pr{\cA} \le \prk{\cA} \le \eu^{+2\alpha_N} \pr{\cA}
\ee

\par\vskip2mm\noindent
$\blacklozenge$
%
It follows from \eqref{eq:prob.BJ.alpha} and \eqref{eq:pr.xi.mu.BK} that
\be
\pr{ \xi\in I_s(\eta) }  \le \sup_{\Bk} \prk{ \xi\in I_s(\eta) }  \le C(F_V,\ell) N\, s.
\ee
Recall that this bound was proved only for $s\le \ell/M(N) = \sord(\ell N^{-1})$.

\qedhere



\begin{bibdiv}
\begin{biblist}



\bib{C10}{misc}{
   author={Chulaevsk{y}, V.},
   title={A remark on charge transfer processes in multi-particle systems},
   status={\texttt{arXiv:math-ph/1005.3387}},
   date={2010},
}


\bib{C11a}{article}{
   author={Chulaevsky, V.},
   title={On resonances in disordered multi-particle systems},
   journal={C.R. Acad. Sci. Paris, Ser. I,},
   volume={350},
   date={2011},
   pages={81--85},
}



\bib{W81}{article}{
   author={Wegner, F.},
   title={Bounds on the density of states in disordered systems},
   journal={Z. Phys. B. Condensed Matter},
   volume={44},
   date={1981},
   pages={9--15},
}


\end{biblist}
\end{bibdiv}
\end{document}